\documentclass[aps,twocolumn,superscriptaddress,showpacs]{revtex4}
\usepackage{amsmath}
\usepackage{graphicx}
\usepackage{graphics}
\usepackage{dcolumn}
\usepackage{bm}
\usepackage{pstricks}

\newcommand{\be}{\begin{equation}}
\newcommand{\ee}{\end{equation}}
\newcommand{\bea}{\begin{eqnarray}}
\newcommand{\eea}{\end{eqnarray}}

\newcommand{\kevr}{{\rm ~keVr}}

\newcommand{\kgday}{kg$\cdot$ day }

\begin{document}

\title{Magnetic Inelastic Dark Matter: 
Directional Signals Without a Directional Detector}

\author{Tongyan Lin}
\affiliation{Physics Department, Harvard University, 
Cambridge, MA 02138, USA}

\author{Douglas P. Finkbeiner}
\affiliation{Physics Department, Harvard University, 
Cambridge, MA 02138, USA}
\affiliation{Harvard-Smithsonian Center for Astrophysics, 
60 Garden St., Cambridge, MA 02138, USA}

\date{\today}

\begin{abstract}

  The magnetic inelastic dark matter (MiDM) model, in which dark
  matter inelastically scatters off nuclei through a magnetic dipole
  interaction, has previously been shown to reconcile the DAMA/LIBRA
  annual modulation signal with null results from other
  experiments. In this work, we explore the unique directional
  detection signature of MiDM. After the dark matter
  scatters into its excited state, it decays with a lifetime of order
  1 $\mu s$ and emits a photon with energy $\sim$100 keV. Both the
  nuclear recoil and the corresponding emitted photon can be detected
  by studying delayed coincidence events. The recoil track and
  velocity of the excited state can be reconstructed from the nuclear
  interaction vertex and the photon event vertex. The angular
  distribution of the WIMP recoil tracks is sharply peaked and
  modulates daily.  It is therefore possible to observe the directional
  modulation of WIMP-nucleon scattering without a large-volume gaseous
  directional detection experiment. Furthermore, current experiments
  such as XENON100 can immediately measure this directional modulation
  and constrain the MiDM parameter space with an exposure of a few
  thousand kg~$\cdot$~day.

\end{abstract}

\pacs{95.35.+d}

\maketitle
\twocolumngrid

\section{Introduction}

Despite decades of direct detection efforts \cite{Gaitskell:2004gd},
the nature of dark matter interactions with regular matter remains
elusive. The results from the DAMA/NaI and DAMA/LIBRA collaborations
suggest that such interactions may be more intricate than originally
expected. DAMA has observed an annual modulation in NaI crystals for
the past decade
\cite{Bernabei:2000qi,Bernabei:2008yi,Bernabei:2010mq}, with the
expected phase for WIMP-nuclei interactions. There is no experimental
evidence corroborating this signal.  By now, it appears that the
signal is not conventional spin-independent elastic scattering of
WIMPs on nuclei.
 
Among the quantitative explanations of DAMA, it is possible to take a
few approaches. One method is to exploit detector effects, such as
channeling \cite{Bernabei:2007hw, Fairbairn:2008gz,
Savage:2008er}. Another is to introduce a dark matter model with more
ingredients (for example, \cite{Ullio:2000bv, Foot:2008nw,
Feldstein:2009tr, Bai:2009cd, Chang:2009yt, Chang:2010yk}). A
possibility is that the dark matter preferentially scatters off the
NaI used in DAMA, as opposed to the nuclei used in other direct
detection experiments. In particular, we focus on the fact that iodine
is special in having both a relatively large mass and a relatively
large magnetic moment \cite{Chang:2010en}.

If dark matter has (weak) electromagnetic moments
\cite{Bagnasco:1993st, Pospelov:2000bq}, it can interact through the
charge and magnetic dipole moment of the nuclei. For a summary of the
interaction strengths for various nuclei used in direct detection
experiments, see \cite{Banks:2010eh}. This type of interaction has been
used to explain some recent direct detection results
\cite{Masso:2009mu,An:2010kc,Cho:2010br,Barger:2010gv,
Fitzpatrick:2010br,Banks:2010eh}, including the positive claim of
DAMA. However, there are strong constraints from CDMS
\cite{Ahmed:2009zw} and XENON \cite{Angle:2009xb, Aprile:2010um} on
this explanation of DAMA.

Inelastic dark matter (iDM) takes advantage of the large iodine mass
\cite{Chang:2008gd}. In iDM, there is an excited state of dark matter
with mass splitting $\delta$. It is assumed that dark matter couplings
with nuclei are primarily off-diagonal, so that the WIMP is scattered
into its excited state.
This interaction only occurs if the dark matter has sufficient
initial velocity. The minimum velocity $v_{min}$ for a WIMP to scatter
with a nuclear recoil of energy $E_R$ is:
\begin{align}
	v_{min} 
	= \frac{1}{\sqrt{2 m_N E_R}} \left( E_R  \left(\frac{m_N}{m_\chi}+1 \right) + \delta \right)
\label{eq:vmin}
\end{align}
where $m_N$ is the nucleus mass and $m_\chi$ is the WIMP mass. For
splittings of $\delta \sim 100$ keV, experiments are only sensitive to
the tail of the WIMP velocity distribution, leading to a much larger
annual modulation than in the elastic case. Furthermore, scattering on
heavier nuclei like iodine is preferred if $m_\chi$ is of order 100
GeV.

The basic iDM model is now tightly constrained
\cite{SchmidtHoberg:2009gn} by the latest results from CRESST
\cite{CRESST:IDM}, ZEPLIN-III \cite{Akimov:2010vk}, XENON
\cite{Angle:2009xb}, and CDMS.
It is possible to combine inelastic scattering with yet
more ingredients. For example, spin-dependent inelastic scattering is
discussed in \cite{Kopp:2009qt}.


\begin{table*}[t]
\begin{center}
\begin{tabular}{|ccc|ccc|ccc|ccc|ccc|ccc|ccc|ccc|}
\hline
& $m_{\chi}$ & & & $\delta$ & & & $\mu_\chi/\mu_N$ & & & 
      $\tau$ & & & $\lambda$ & & & $\eta_{.15}$ & & & Angular Rate  & & & 
      XENON100 & \\
& (GeV) & & & (keV) & & &  & & & ($\mu$s) & & & (m) & & &  
  & & & $10^{-3}$(cpd/kg)  & & & (non-blind) & \\
\hline
& 70* & & & 123 & & & $6.2\times 10^{-3}$ & & & 1.2 & & & 0.4 & & & 0.23 & & & 11.3 & & & 1.4 & \\
& 140* & & & 109 & & & $2.2\times 10^{-3}$ & & & 12.7 & & & 6.2 & & & 0.018 & & & 2.2 & & & 8.1 &\\
& 300* & & & 103 & & & $2.0\times 10^{-3}$ & & & 18.0 & & & 9.7 & & & 0.012 & & & 1.7 & & & 11.6 &\\
\hline
& 70 & & & 135 & & & $11.2\times 10^{-3}$ & & & 0.26 & & & 0.09 & & & 0.63 & & & 17.6 & & & 0.07 & \\
& 140 & & & 125 & & & $3.2\times 10^{-3}$ & & & 3.9 & & & 2.0 & & & 0.06 & & & 4.4 & & & 3.3 & \\
& 300 & & & 117 & & & $2.5\times 10^{-3}$ & & & 7.9 & & & 4.4 & & & 0.03 & & & 2.6 & & & 5.8 & \\
\hline
& 70 & & & 100 & & & $2.5\times 10^{-3}$ & & & 12.6 & & & 4.9 & & & 0.024 & & & 2.7 & & & 9.2 &\\
& 140 & & & 90 & & & $1.6\times 10^{-3}$ & & & 42.2 & & & 20.2 & & & 0.006 & & & 1.3 & & & 22.2 &\\
& 300 & & & 90 & & & $1.6\times 10^{-3}$ & & & 42.2 & & & 22.1 & & & 0.005 & & & 1.0 & & & 19.3 & \\
\hline
\end{tabular}
\end{center} 
\caption{In the first three (starred) rows, we give the best fit
benchmark models of MiDM, with $v_{esc} = 550$ km/s and $v_0 = 220$
km/s \cite{Chang:2010en}. We also list parameters within the 90$\%$ CL
region of each best fit value, for which the lifetime, $\tau$, can be
a factor of a few larger or smaller. $\lambda$ is the average recoil
track length. $\eta_{.15}$ is an estimate of the efficiency of
XENON100 to detect delayed coincidence events, as described in
Section~\ref{sec:detectoreff}. The `angular' rate is the rate for
delayed coincidence events with a nuclear recoil in the energy range
$10-80$ keVr, followed by a photon with $\delta$ keVee. This is
obtained from multiplying the total rate by $\eta_{.15}$. We also show
the expected number of nuclear recoil events for the published
XENON100 non-blind analysis.}
\label{tab:benchmarks}
\end{table*}


\subsection{Magnetic Inelastic Dark Matter}

We focus on magnetic inelastic dark matter (MiDM), because it has a
unique and interesting directional signature. Chang et
al. \cite{Chang:2010en} showed MiDM could explain both DAMA and other
null results. The model takes advantage of both the magnetic moment
and large mass of iodine. In MiDM, the dark matter couples
off-diagonally to the photon:
\begin{equation}
  {\mathcal L} \supset \left( \frac{\mu_\chi}{2} \right) \overline \chi^* \sigma_{\mu \nu} F^{\mu \nu} \chi + c.c.
\end{equation}
where the mass of $\chi$ and $\chi^*$ are split by $\delta \sim 100$
keV. The off-diagonal coupling is natural if the dark matter is a
Majorana fermion. The excited state has a lifetime $\tau
= \pi/(\delta^3 \mu_\chi^2) \sim 1-10 \mu$s, and emits a photon when
it decays. This short lifetime makes it possible to observe both the
nuclear recoil and the emitted photon with a meter-scale detector. The
two interaction vertices allow reconstruction of the excited state
track. Both the velocity and angle can be measured, enabling
directional detection even without a directional detector.

A dark matter particle with a permanent electromagnetic dipole moment
generally can be constrained by, e.g., gamma-ray measurements, the
CMB, or precision Standard Model tests \cite{Goodman:2010qn,
Sigurdson:2004zp, Gardner:2008yn}. However, the strongest bounds tend
to come from direct detection experiments themselves, at least in the
100 GeV mass range. Furthermore, in MiDM, the inelastic nature of the
interaction suppresses interactions with photons and baryons at low
energies. If the dark matter is a composite particle, a low
compositeness scale can also suppress annihilation to photons.

There are some variants of the MiDM idea. In \cite{Feldstein:2010su},
the parameter values were taken to be $m_\chi \sim 1$ GeV and $\delta
\sim 3$keV. The DAMA signal is produced by the emitted photon.
This explanation evades constraints
from other direct detection experiments because such low-energy
electromagnetic events are typically rejected or not seen by other
detectors. 

It is also possible that the dark matter couples to a new `dark'
$U(1)$, with gauge boson mass $m_A \neq 0$
\cite{Holdom:1986eq,Pospelov:2007mp,An:2010kc}.  Here the dark matter
has a large dark dipole. If the dark gauge boson couples to
regular electromagnetic currents, a sizable interaction with nuclei
can be generated. However, the decay rate of the excited state is
suppressed because there is no direct interaction with the
photon. While these ideas are interesting explanations of the DAMA
signals, we do not consider them further because the excited state has
a long lifetime.

We study MiDM benchmarks, given in Table~\ref{tab:benchmarks}, which
are good fits to the DAMA annual modulation signal
\cite{Chang:2010en}. MiDM models with $m_\chi$ greater than $\sim 300$
GeV are severely constrained by ZEPLIN-III \cite{Akimov:2010vk}, KIMS
\cite{Kim:2008zzn}, and XENON100 \cite{Aprile:2010um}.

The benchmarks are subject to form factor and velocity distribution
uncertainties
\cite{MarchRussell:2008dy,Kuhlen:2009vh,Ling:2009eh,Vogelsberger:2008qb,
Alves:2010pt}, especially for larger masses. The directional signal
prediction can change wildly depending on the lifetime and rate.

In order to explore the parameter space, we also considered two
extreme points within the DAMA 90$\%$ confidence level region found by
\cite{Chang:2010en}, for each of the three masses. For the point with
highest $\delta$ and $\mu_\chi$, the expectation for directional
detection is better. The point with lowest $\delta$ and $\mu_\chi$,
which would not result in many delayed coincidence events, is in any
case already tightly constrained by the XENON100 non-blind analysis.

In this paper, we show that the current generation of direct detection
experiments can observe a directional signal from MiDM. For
concreteness we focus on a XENON100-like detector, for two reasons.
First, XENON100 will soon place strong constraints on the MiDM
parameter space, making it the most relevant experiment to
consider. Second, we wish to emphasize the feasibility of detecting a
directional signal with experiments that are currently running.

We compute the distribution of recoil track angles and velocities from
MiDM benchmarks. The sensitivity of XENON100 to the MiDM parameter
space depends strongly on the lifetime of the excited state. For the
benchmark lifetimes of $\sim 1-10 \mu$s, XENON100 can measure the
directional modulation at high significance and obtain sharp
constraints on the parameter space with just tens of events. This is
achievable with around 5000 kg~$\cdot$~day in the energy range $10-80$
keVr.

\section{Directional Detection}

Directional detection can clearly test whether any signal comes from
WIMP interactions \cite{Spergel:1987kx}. Due to the Earth's motion in
the Galaxy, there is a ``WIMP wind'' which is opposite the motion of
the Earth.  There is a daily modulation in the angle of recoil tracks
in the lab frame. This modulation depends only on the rotation of the
Earth relative to the WIMP wind, and can be disentangled from the
daily rotation of the Earth with respect to the Sun. The experimental
directional detection effort focuses on measuring the nuclear recoil
track with large-volume, gaseous detectors \cite{Ahlen:2009ev,
Sciolla:2008vp, Sciolla:2009ps}.


Angular information is a particularly powerful discriminant of WIMP
scattering for iDM \cite{Finkbeiner:2009ug,Lisanti:2009vy}. Because
inelastic interactions have a high velocity threshold, the angular
distribution of the nuclear recoil tracks is sharply peaked in the
direction of the WIMP wind. There is a kinematic constraint on the
recoil angle of the nucleus:
\begin{equation}
  (\cos \gamma)_{max}(E_R) = \frac{v_{esc} - v_{min}(E_R,\delta)}{v_E}.
  \label{eq:cosgamma}
\end{equation}
Here $\gamma$ is the angle between the velocity of the Earth and the
recoil velocity in the Earth frame, $v_E$ is the Earth's velocity in
the Galactic frame, and $v_{esc}$ is the Galactic escape velocity from
the Solar neighborhood. 
For typical iDM models considered in the literature, $\gamma$ is
constrained to be within $\sim$100 degrees of the WIMP wind
\cite{Finkbeiner:2009ug}. However, because the signal goes to zero at
the bound in Eq.~\ref{eq:cosgamma}, the precise location of this
kinematic constraint can be difficult to pinpoint.

MiDM has better directional detection prospects at XENON100, compared
to directional detection of iDM. Current directional detectors focus
on spin-dependent scattering and use light targets such as CF$_4$
\cite{Ahlen:2010ub, Daw:2010ud, Grignon:2010zg, Miuchi:2007ga}. Thus,
they would not see inelastic scattering events.  In the MiDM case,
there is also much more event information and thus more sensitivity to
the parameter space. One can measure both the velocity ($v_*$) and the
angle ($\cos \gamma_*$) of the WIMP recoil track. Once again, this
recoil angle is with respect to the Earth's motion. The tracks are
sharply peaked in angle opposite the motion of the Earth.

For the WIMP recoil angle, there is also an energy-dependent maximum
recoil angle, which we give in Sec.~\ref{sec:spectrum}. The most
important bound is on the WIMP recoil velocity,
\begin{equation}
	v_*^{min}(E_R) = \left| \frac{(E_R (m_N/m_\chi -1) - \delta)}{\sqrt{2 m_N E_R}} \right|.
	\label{eq:vfmin}
\end{equation}
Here the signal peaks near the kinematic bound because most events
occur near the threshold velocity in Eq.~\ref{eq:vmin}. Thus having
information on both $v_*$ and $E_R$ is an extremely sensitive probe of
the model parameters. There is a remaining degeneracy: if $\delta$ and
$m_\chi$ are shifted in {\emph{opposite}} directions, the bound can
remain roughly the same. However, one can fit $\delta$ separately from
the spectrum of the nuclear recoils, and from the energy of the
emitted photons.

There is also a maximum velocity for the excited state,
\begin{equation}
	v_*^{max}(E_R) =\sqrt{(v_E + v_{esc})^2 -2 (E_R + \delta)/m_\chi},
\end{equation} 
but the rate is exponentially suppressed at this bound.

\subsection{XENON100 \label{sec:xenon100}}

We model directional detection in XENON100 with a simplified
XENON100-like experiment. XENON10 \cite{Angle:2007uj,Angle:2009xb} had
316.4 \kgday of data in the energy range 4.5-75 keVr. XENON100 has a
40 kg fiducial mass, at even lower backgrounds. The initial 170 \kgday
non-blind run already constrains the MiDM parameter space (at low
$\delta$). 

The XENON100 detector is a cylinder, with a radius of 15.3 cm and a
height of 30.6 cm. The fiducial volume has a radius of 13.5 cm and
height of 24.3 cm. The primary scintillation (S1) and ionization (S2) signals
of an event are measured. For more details,
see \cite{Aprile:2010bt}. The S2 signal is observed 15-140 $\mu s$
after the S1 signal, for events in the fiducial volume.

The signature of MiDM is two S1 signals separated by roughly .5 $\mu
s$ in time, followed at least 15 $\mu s$ later by two S2 signals. The
photon event is identified from the second S1 signal and an S2 peak
with energy of $\sim$100~keVee. At 100~keVee, a photon is clearly
distinguishable from a nuclear recoil by S2/S1. The other event should
be consistent with a nuclear recoil. The time separation of the two S2
signals depends on how the WIMP recoils along the cylinder axis,
$z$. In XENON10, events with multiple S2 events at different $z$
positions were rejected.

We refer to the track connecting the two events as the decay track.
Events can be localized to a 3D spatial resolution of 3 mm (though the
absorption length for the 100 keV photon may blur this) and timing
resolution of 10 ns. Meanwhile, the track should be at least 10 cm
long.  This makes it possible to measure direction and velocity of the
decay track to an extremely high accuracy.  The head-tail
discrimination of the track can be determined using timing information
and the S1/S2 ratio.

We wish to obtain the $\chi^*$ recoil track from the decay
track. However, because the photon can travel up to $\sim$1 cm after
emission, this introduces systematic uncertainties.  The observed
decay track can be blurred by a few degrees, relative to the $\chi^*$
recoil track direction.  This also introduces an uncertainty in the
velocity of the $\chi^*$ of roughly 10\%.

There are some specific event geometries that can result in more
ambiguous events. For example, it could be difficult to resolve the
two S2 signals if the decay track is perpendicular to the $z$
axis. Then the two S2 signals arrive at nearly the same time. S2
signals generally have a time width of $\sim1 \mu$s and the PMT
spatial resolution is only $\sim$2.5 cm. However, because the drift
velocity is 2mm/$\mu$s, this is a small fraction of the total solid
angle.

Thus directional events are in principle detectable at XENON100. The
background for such delayed coincidence events with both a nuclear
recoil and a photon of energy $\sim 100$ keV should be extremely
low. There are other `mixed' delayed coincidence events from Bi and Kr
contamination, and excitation of metastable states of Xe
\cite{Aprile:2010bt}. However, these have very different energies and
decay times. It may be possible to extend the fiducial volume when
searching for directional events.

\begin{figure}[t]
\centering
\includegraphics[width=.5\textwidth]{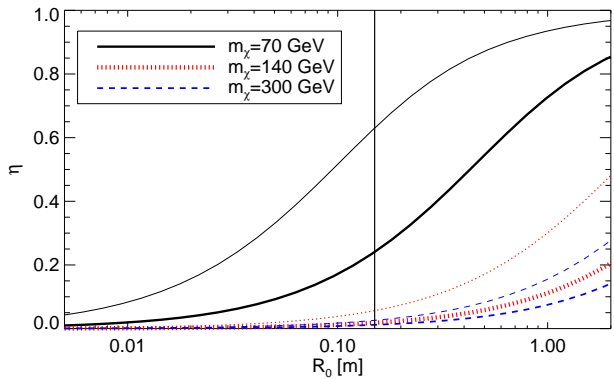} 
\centering
\caption{The efficiency $\eta$ for the best-fit benchmarks from
Table~\ref{tab:benchmarks}. $R_0$ is the size of a spherical
detector. We approximate the XENON100 fiducial volume as a sphere with
radius $R_0 = 0.15$ m, marked by the vertical black line. The thinner
lines show the corresponding results with highest $\delta$, within the
90$\%$ CL region of the best fit.}
\label{fig:eff}
\end{figure}

\subsection{Detector Efficiency \label{sec:detectoreff}}

The typical decay length is $1-10$ m in these models, relatively large
compared to XENON100. Thus the WIMP can recoil inside the detector
volume, but decay outside the detector
\footnote{ The reverse can also happen, similar to the idea in
\cite{Feldstein:2010su}. The rate depends on whether the material
within $\sim$10m of the Xe detector mostly consists of light or heavy
nuclei. Aside from a 20cm layer of lead, the shielding for XENON100
consists of polyethylene, water, and copper.}.  The effective exposure
for delayed coincidence events is, in general, lower than the exposure
for nuclear recoils because of this geometric effect. Here we compute
the detector efficiency, as a function of typical detector size, for
the MiDM benchmarks.

The efficiency is
\begin{equation}
   \eta(t) =  \int d^3 \vec v_* f(\vec v_*,t)
      \int \frac{dt'}{\tau} e^{-t'/\tau} 
      \left( \int_{V} \frac{d^3 \vec x}{V} 
      H(t,\vec x, \vec d,t') \right) \nonumber
\end{equation}
The term in brackets comprises detector effects. The spatial integral
is over the detector volume. $H(t, \vec x, \vec d, t')$ is the
efficiency for observing a WIMP decay, given that a nuclear recoil was
observed. This depends on the time of the year $t$, the location of
the WIMP-nucleus interaction inside the detector, $\vec x$, the decay
vector, $\vec d$, and the WIMP decay time (coincidence time), $t'$.
Whether a given WIMP decay track is located inside the detector
depends on the orientation of the detector with respect to the Earth's
velocity, the decay vector, and the efficiency for the particular
event geometry.

The astrophysics and particle physics is captured by the integral over
$t'$ and $\vec v_*$. $\tau$ is the lifetime of the excited state. The
distribution of recoils depends on the WIMP recoil velocity
distribution, $f(\vec v_*)$, and the decay time distribution. We
assume that $\vec v_*$ is defined with respect to the Earth's velocity
vector so that $f(\vec v_*)$ does not depend on detector orientation.

For the calculation below, we model the detector as a single sphere of
size $R_0$. We assume that $H$ depends only on the interaction
position $\vec x$ and the decay length $L = v_* t'$. Here we neglect
the smearing arising from the mean free path of the emitted photon,
since the emission is isotropic. There is also no dependence on $t$ or
recoil angle in this approximation. Then the expression for efficiency
above can be simplified to
\begin{equation}
   \eta = \int dL\ g(L)\ \int_0^{R_0} \frac{3 R^2 dR}{R_0^3} H(R,L)
\end{equation} where $L$ is the recoil length. The recoil length
distribution $g(L)$ is
\begin{equation}
   g(L) = \int dv_* \frac{f(v_*)}{v_* \tau} \exp\left( -\frac{L}{v_* \tau} 
    \right)
\end{equation}
where now $f(v_*)$ is the distribution for $v_*$, not $\vec v_*$. A
good approximation is $g(L) = \exp( -L/\lambda)/\lambda$, where
$\lambda = \langle v_* \rangle \tau$ is the average recoil
length. Typical $\lambda$ values are given in
Table~\ref{tab:benchmarks}.

We approximate the XENON100 detector as a sphere. The fiducial volume
has radius $R_0 = 15$ cm, with efficiency $\eta_{.15}$.  Results are
shown in Fig.~\ref{fig:eff}. The precise efficiency depends on
specifics of the detector, and must take into account the effects
mentioned in Sec.~\ref{sec:xenon100}.

\begin{figure*}[ht]
\begin{center}
a)\includegraphics[width=.32\textwidth]{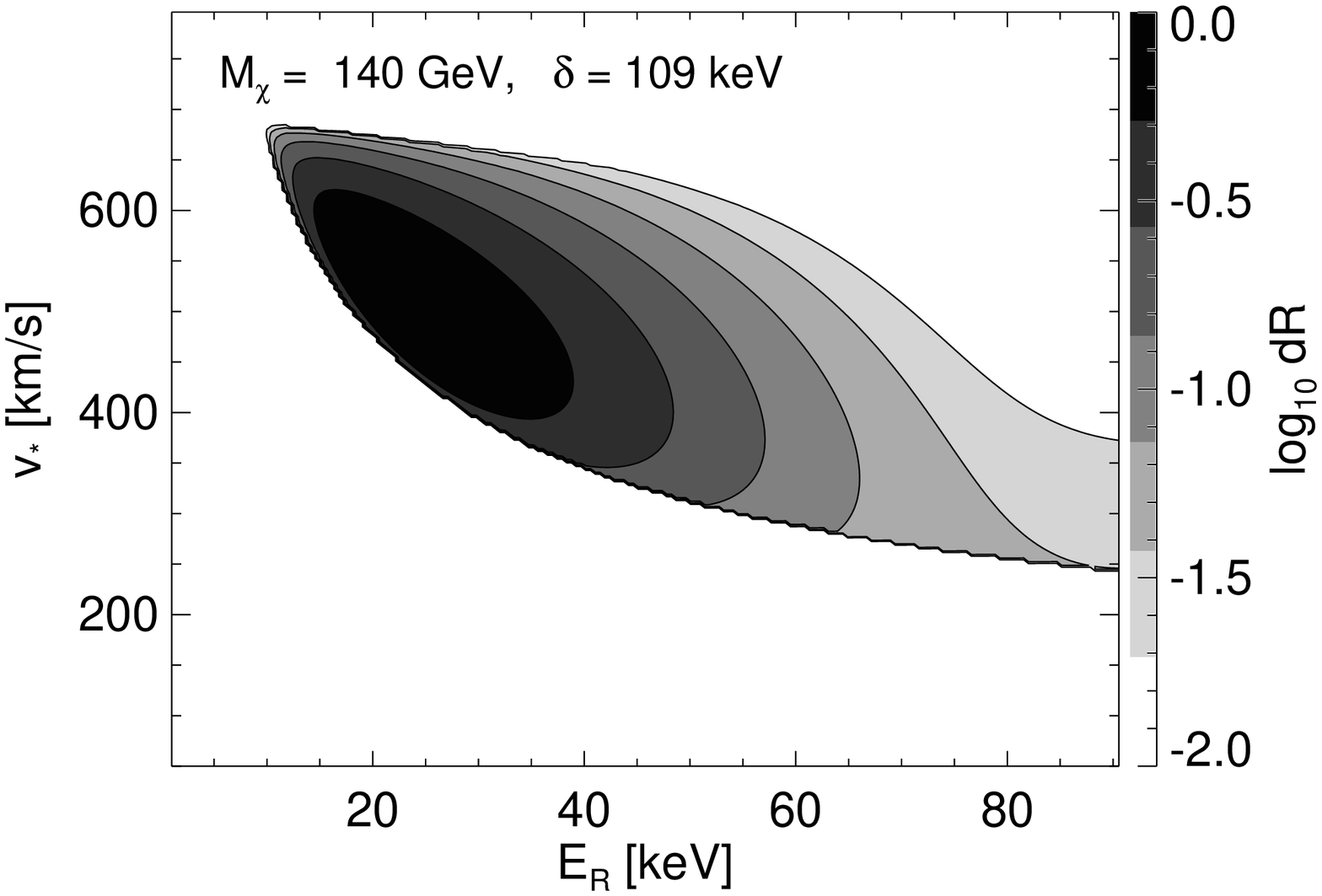}
\includegraphics[width=.32\textwidth]{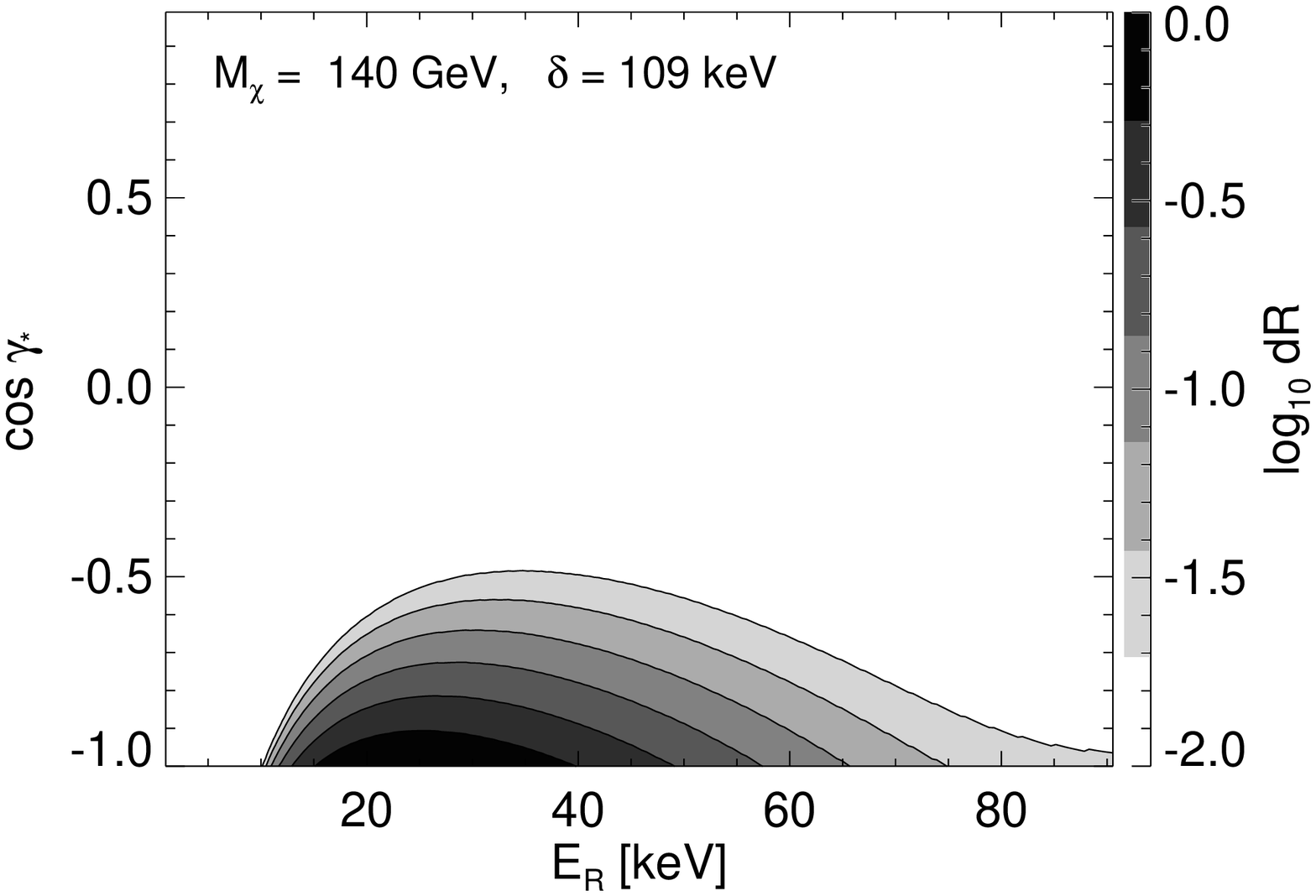}
\includegraphics[width=.32\textwidth]{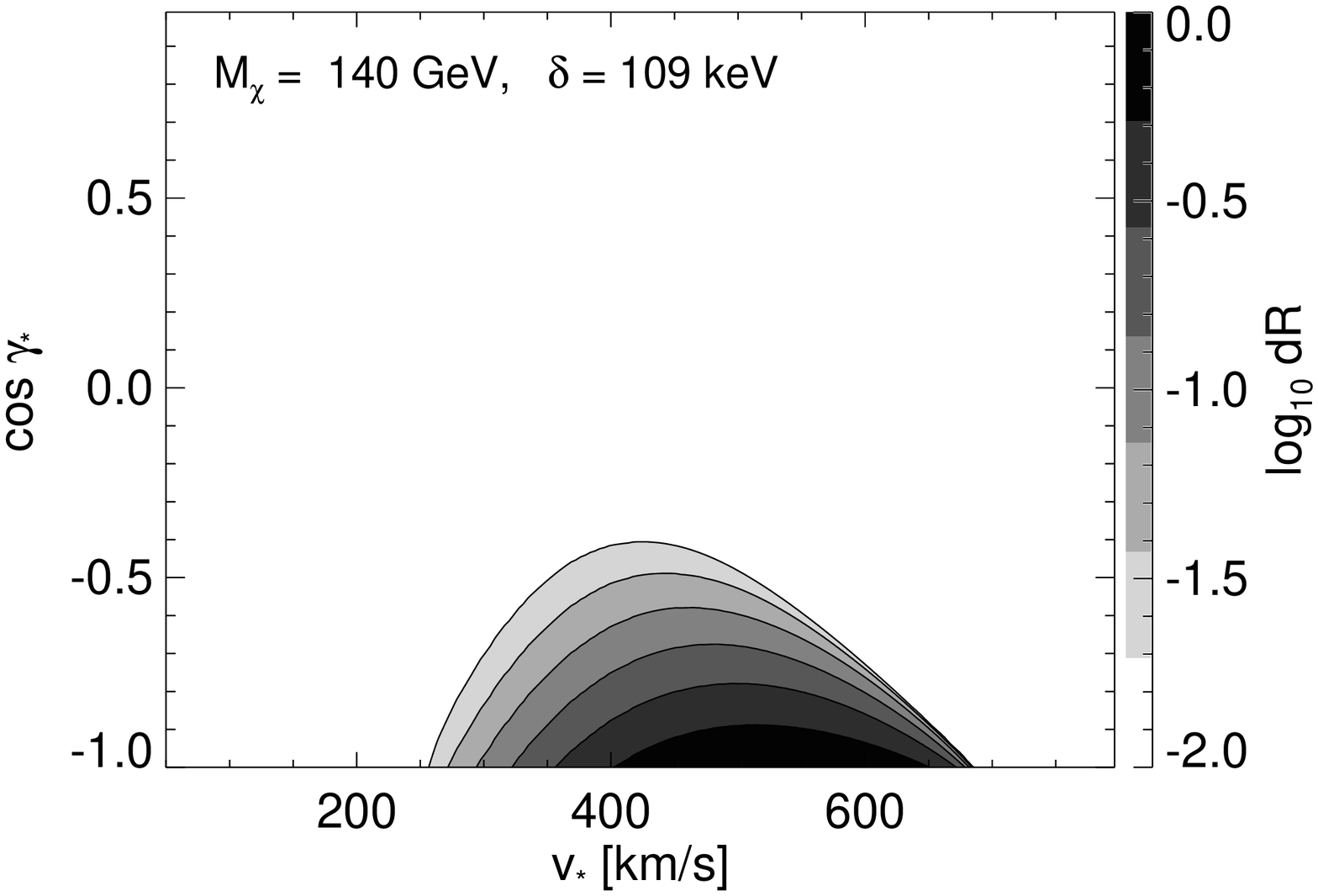}\\
b)\includegraphics[width=.32\textwidth]{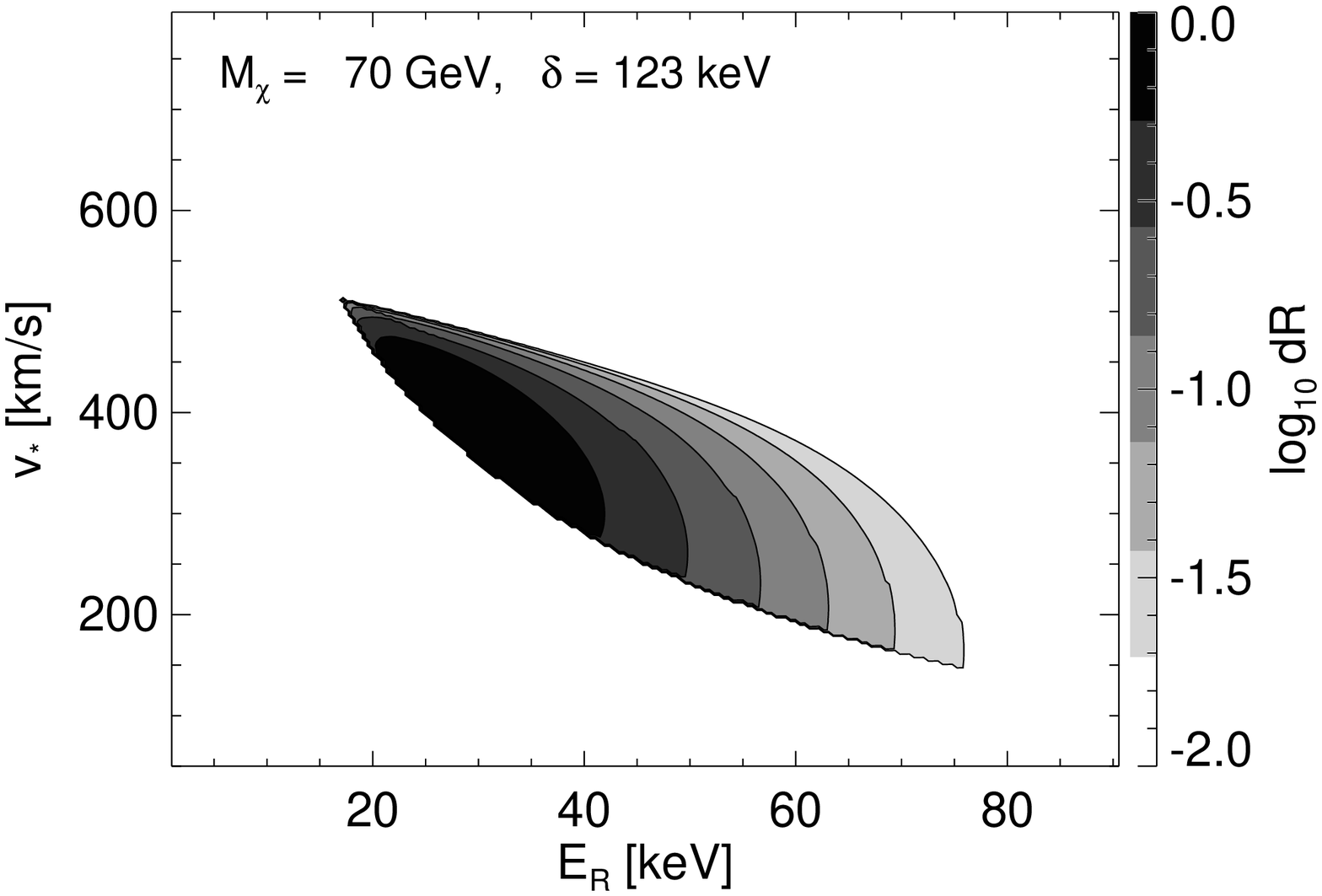}
\includegraphics[width=.32\textwidth]{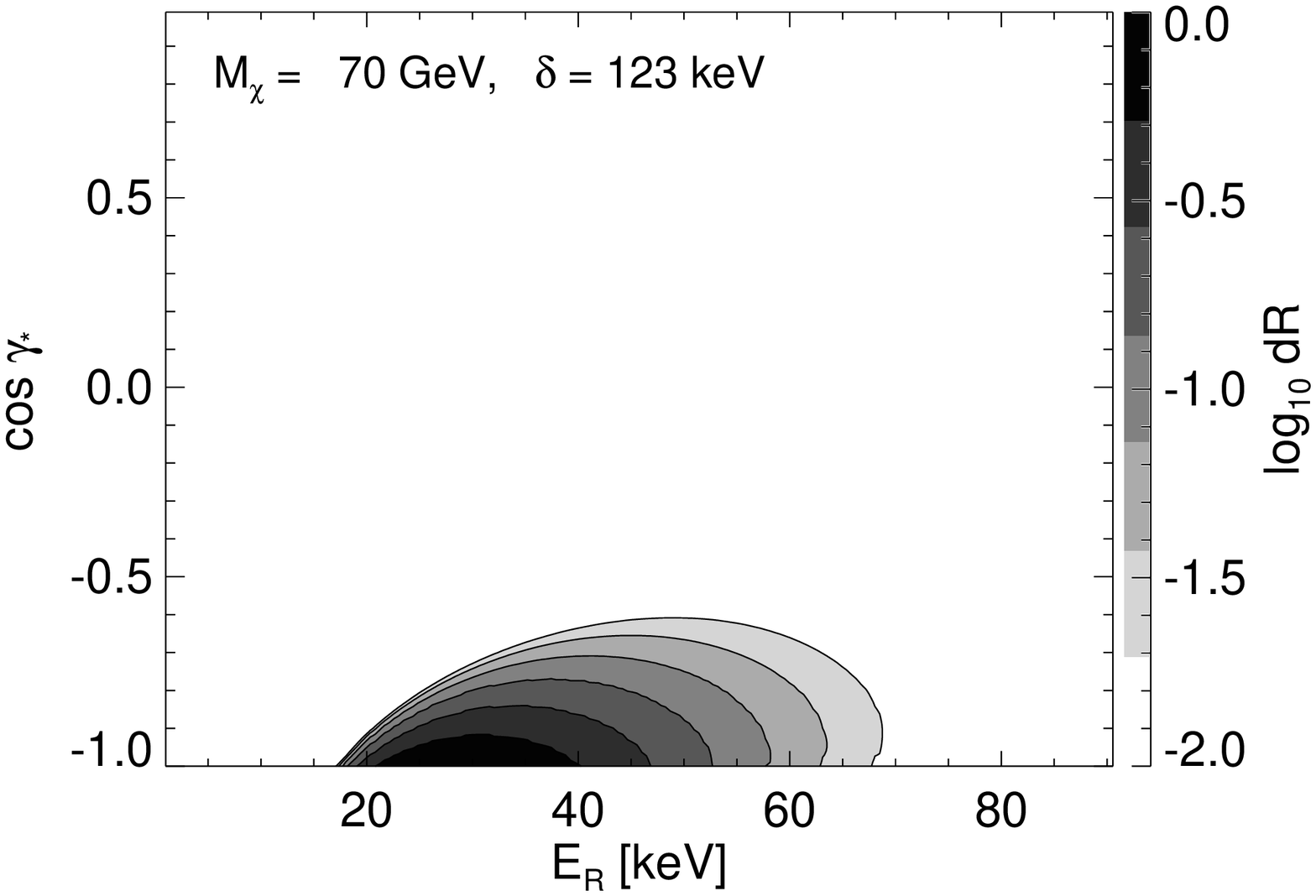}
\includegraphics[width=.32\textwidth]{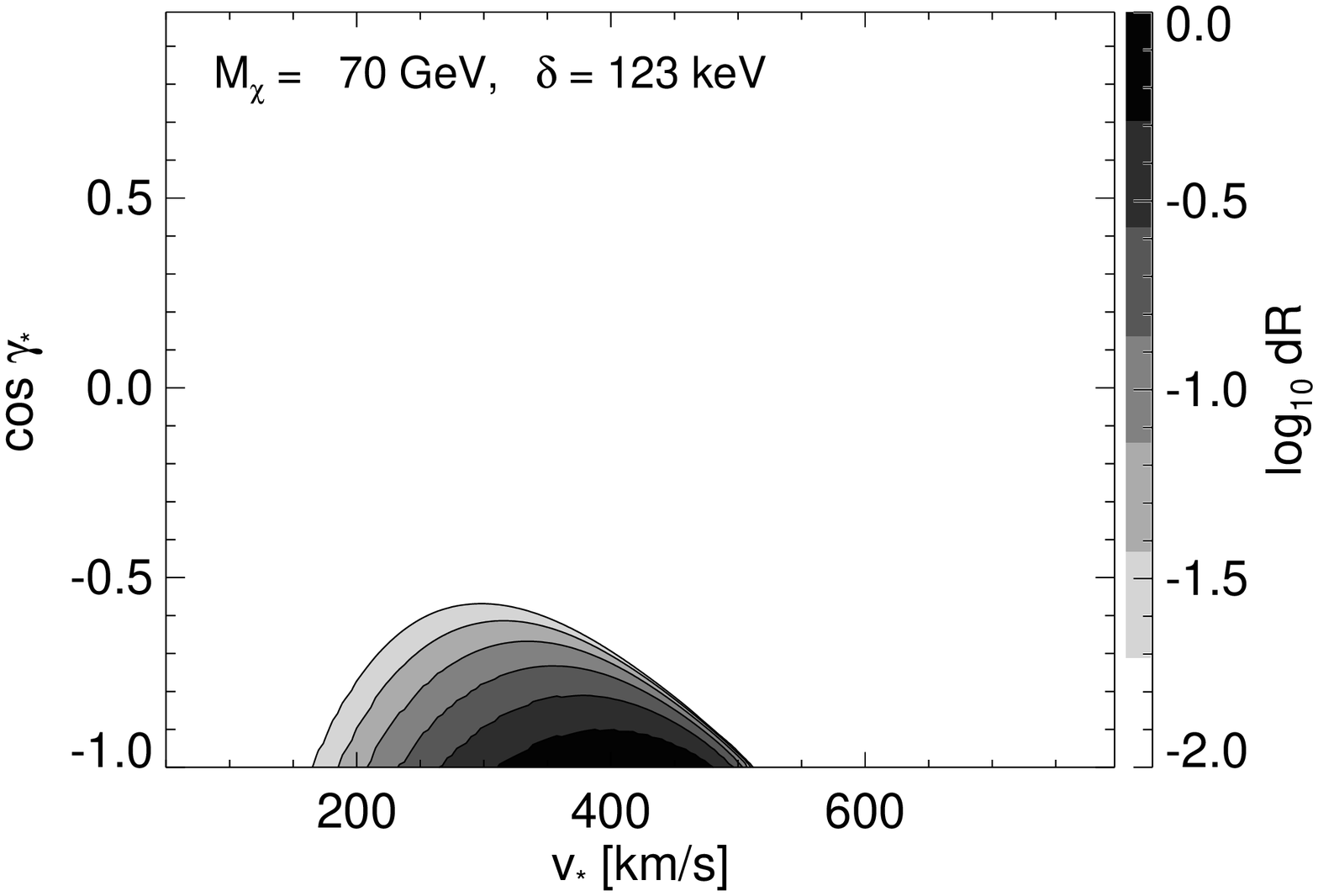}\\
\end{center}
\caption{Differential rates $dR/(dE_R d\cos\gamma_* dv_*)$ for (a)
  $m_\chi = 140$ GeV MiDM benchmark and (b) $m_\chi = 70$ GeV MiDM
  benchmark. The $m_\chi = 300$ GeV benchmark looks similar to the
  $m_\chi = 140$ GeV benchmark. In each case we show three
  two-dimensional distributions, where we have integrated over the
  third variable. All rates are computed assuming scattering on Xe,
  and benchmarks are given in Table~\ref{tab:benchmarks}. The
  differential rate is normalized so that the total rate is unity.}
\label{fig:rates}
\end{figure*}

\section{Recoil Spectrum \label{sec:spectrum}}

There are two electromagnetic scattering channels for
magnetic dark matter: dipole-dipole and dipole-charge.  In the
dipole-dipole scattering case, the dark matter interacts with the
magnetic moment of the nucleus. The matrix element is
\begin{equation}
	\frac{|{\cal M}|^2}{32 \pi m_N  m^2_\chi} = 16 \pi \alpha^2 m_N
 \left( \frac{\mu_{nuc}}{e}\right)^2 
 \left( \frac{ \mu_\chi }{e}\right)^2 \frac{S_N+1}{3 S_N},
 	\label{eq:matrixelementDD}
\end{equation}
for each isotope. We sum over all isotopes, weighted by their
abundances \cite{Banks:2010eh}. There is, in general, also a factor of
$(S_\chi + 1)/(3 S_\chi)$ for the spin of the dark matter. We take
$S_\chi = 1/2$.

In the dipole-charge scattering case, the dark matter interacts with
the electric charge of the nucleus. The matrix element is
\begin{align}
\frac{|{\cal M}|^2}{32 \pi m_N  m^2_\chi} =& \frac{4 \pi Z^2 \alpha^2}{E_R} 
	 \left( \frac{\mu_\chi }{e}\right)^2 
 \big[v^2 - E_R\left( \frac{1}{2 m_N} + \frac{1}{m_\chi} \right) \nonumber \\
 &-\delta \left( \frac{1}{\mu_{N\chi}} +\frac{\delta}{2 m_N E_R} \right) \big],
 \label{eq:matrixelementDZ}
\end{align}
where $v$ is the initial velocity of the WIMP in the lab frame. We
have again assumed $S_\chi = 1/2$.

The differential scattering rate for measuring both nuclear recoil
energy and WIMP recoil track is
\begin{equation}
	\frac{dR}{dE_R dv_* dx_*} =
        \frac{\eta N_T \rho_\chi}{m_\chi}  \int d^3\vec v 
        \ f(\vec v + \vec v_E)\ v \ \frac{ d\sigma }{dE_R dv_* dx_*}
\end{equation}
where we have abbreviated $x_* = \cos \gamma_*$.  The
three-dimensional WIMP velocity distribution is given by $f(\vec v)$.
$N_T$ is the number of target nuclei per kg and $\rho_\chi$ is the
local WIMP energy density, which we fix to be $0.4$ GeV/cm$^3$
\cite{Catena:2009mf}.

As in \cite{Finkbeiner:2009ug}, we expand $d \sigma$ and change
variables to $\vec v' = \vec v + \vec v_E$. The trivial integral over
$\vec v'$ imposes the condition
\begin{equation}
  \vec v' = \vec q/m_\chi + \vec v_* + \vec v_E.
\end{equation} 
$\vec q$ is the recoil momentum of the nucleus.  The resulting
differential rate is
\begin{align}
  dR =
   \frac{\eta N_T \rho_\chi}{m_\chi} & d^3 \vec v_* 
    d^3 \vec q \ f(\vec v')
   \left( \frac{|{\cal M}|^2}{64 \pi^2 m_\chi^2 m^2_N}  \right) F^2[E_R] 
   \nonumber \\
    & \times \delta^{(1)} 
 \left( \frac{q^2}{2 m_\chi} + \vec q \cdot \vec v_* - E_R - \delta \right).
\end{align}
$F^2[E_R]$ is a nuclear form factor which depends on the type of
interaction.

For a xenon target, dipole-charge scattering,
Eq.~\ref{eq:matrixelementDZ}, dominates. For this we use the standard
nuclear Helm form factor. Dipole-dipole scattering,
Eq.~\ref{eq:matrixelementDD}, is roughly $20\%$ of the total rate.  To
calculate dipole-dipole scattering a magnetic moment form factor is
necessary. The nuclear magnetic moment receives contributions from
both spin and angular momentum. We use the spin form factor from
\cite{Ressell:1997kx}.  The angular momentum component is $\sim
20-30\%$ at zero momentum for Xe. Since dipole-dipole scattering is
already subdominant for Xe, and since we do not have accurate angular
momentum form factors, we approximate the entire magnetic moment form
factor with the spin component.

We now specialize to the case where $f(\vec v)$ is a normalized,
truncated Maxwell-Boltzmann distribution, with $v_{esc} = 550$ km/s
\cite{Smith:2006ym} and $v_0 = 220$ km/s. We assume $v_E = 240$ km/s
on average and label the normalization factor of the distribution as
$n(v_0,v_{esc})$. The result is
\begin{align}
  \frac{dR}{dE_R dv_* dx_*} =\ &
\frac{\eta N_T \rho_\chi v_*}{m_\chi} 
   \frac{|{\cal M}|^2}{32 \pi m_N m_\chi^2}
   F^2[E_R] \Theta(1-|x_q|) \nonumber \\
 & \times \int d\phi 
  \frac{e^{ -(v ')^2/v_0^2}}{n(v_0,v_{esc})} \Theta(v_{esc}-|\vec v'|) 
\end{align}
with the following definitions:
\begin{align}
 x_q =& -\frac{(E_R (m_N/m_\chi -1) - \delta)}{q v_*}, \ \ \ \text{and} \\
 (v')^2 =& v_E^2 +q^2/m_\chi^2 + v_*^2 + 2 v_E v_* x_* + 2  x_q v_* q / m_\chi 
 \nonumber \\	
 & + 2  v_E q /m_\chi \left( x_q x_* + \sqrt{1-x_q^2} \sqrt{1-x_*^2}
 \cos \phi \right).  \nonumber
\end{align}
An upper bound on $x_*$ can be extracted from setting $v' = v_{esc}$,
with $\cos \phi = -1$. The bound depends on both $v_*$ and $E_R$.

Finally, the matrix elements are given in Eq.~\ref{eq:matrixelementDD}
or in Eq.~\ref{eq:matrixelementDZ}.  Note that in the dipole-charge
scattering case we need to replace $v$ in Eq.~\ref{eq:matrixelementDZ}
using the energy conservation relation, $ m_\chi v^2 = 2 E_R + m_\chi
v_*^2 + 2 \delta$.

The normalized total rate spectrum of several benchmark models is
shown in Fig.~\ref{fig:rates}.

\begin{figure*}[thb]
\centering
a) \includegraphics[width=.45\textwidth]{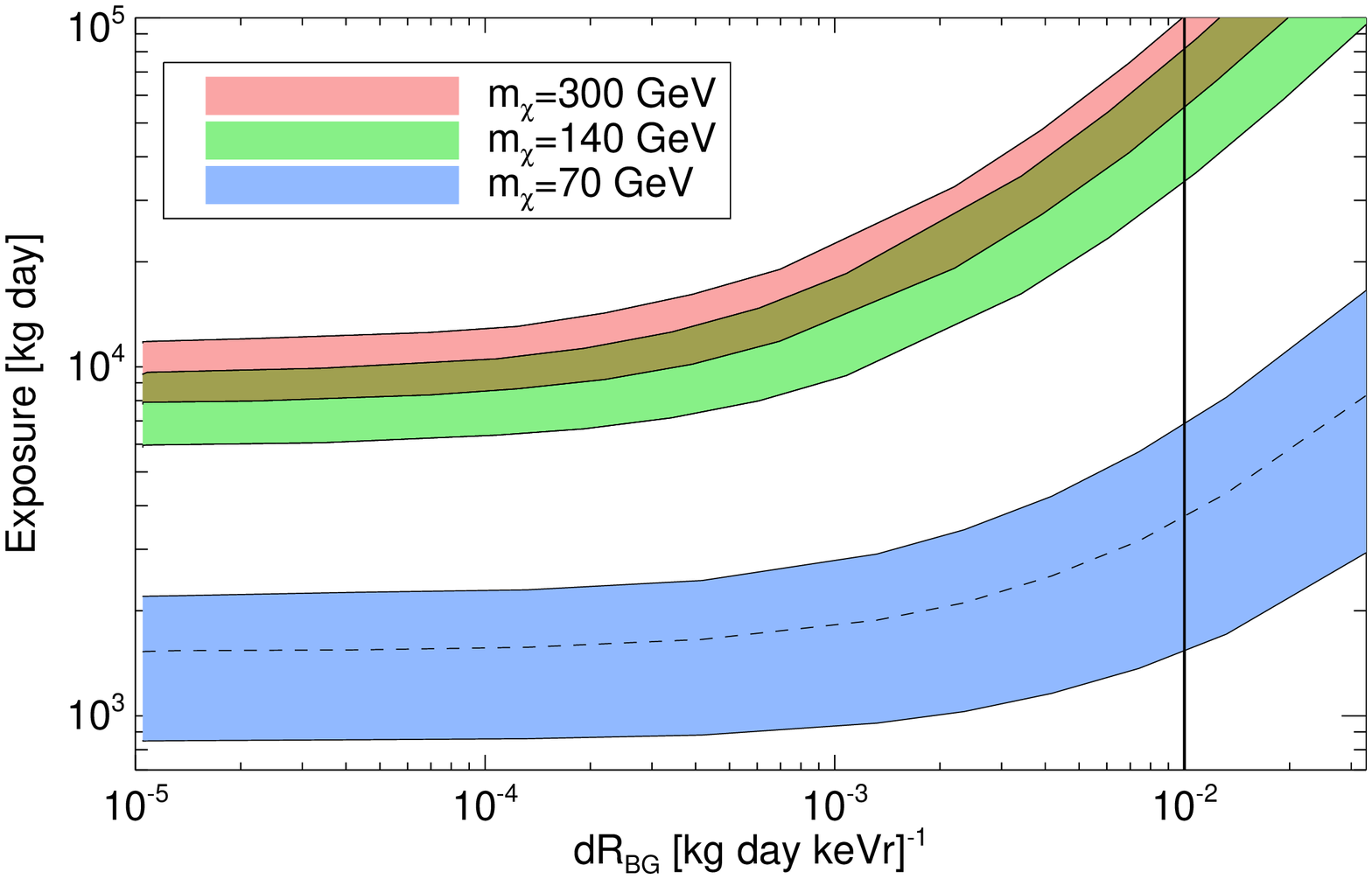}
b) \includegraphics[width=.45\textwidth]{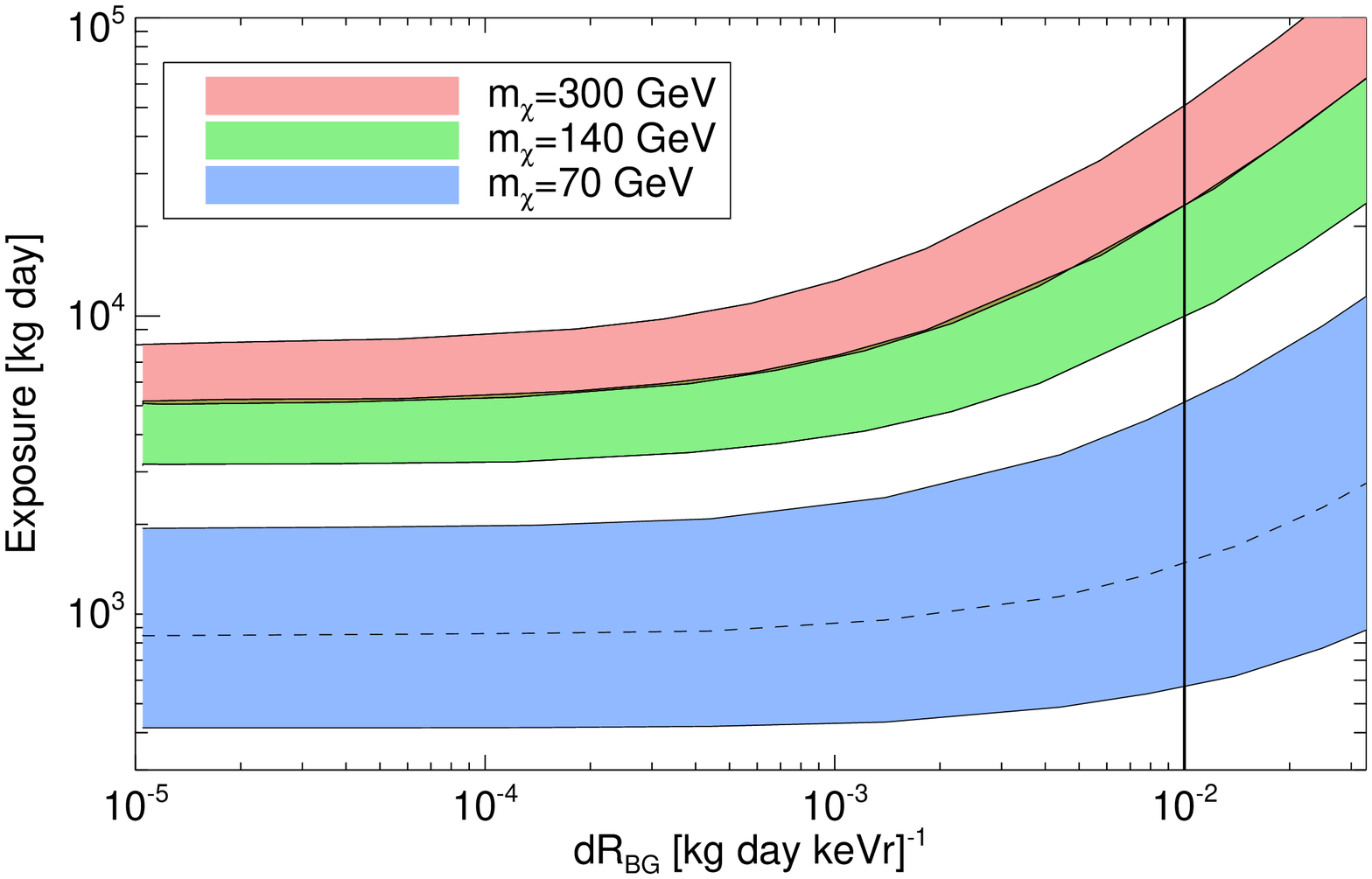} 
\centering
\caption{Exposure to obtain a 5$\sigma$ measurement of $\langle \cos
\gamma_* \rangle$ 90$\%$ of the time the experiment is conducted on
Earth. The energy range of the experiment is $10-80$ \kevr. $dR_{BG}$
is the background rate; a XENON maximum background rate is indicated
by the solid vertical line. The bands shown give the exposures
necessary as the rates modulate throughout a year. We show (a) three
mass benchmarks and (b) three mass benchmarks in the 90$\%$ CL region
with highest delta, from Table~\ref{tab:benchmarks}.}
\label{fig:expbench}
\end{figure*}

\section{Sensitivity}

XENON100 is collecting several thousand \kgday of
exposure. We assume a total exposure of 5000 \kgday on a 40.6
kg fiducial target, in a nuclear recoil energy range of 10-80
$\kevr$. This is consistent with scaling up the results from XENON10
and with preliminary results reported by XENON100.

For the best-fit parameters listed in Table~\ref{tab:benchmarks}, this
would imply a minimum of $\sim 100$ nuclear recoils observable by
XENON100.  Only $\sim 10$ delayed coincidence events are expected, due
to the small size of the detector relative to the average recoil track
length.  Despite these low efficiencies, a study of the delayed
coincidence events is still vastly more informative in two ways: (a)
it establishes a directional signal correlated with the WIMP wind, and
(b) it is much more sensitive to the parameter space.


\subsection{Directional Detection \label{sec:cosgamma}}

We first determine the exposures required to establish a correlation
with the WIMP wind. The average nuclear recoil angle with respect to
the Earth's motion, $\langle \cos \gamma \rangle$, is a robust
model-independent statistic for directional detection
\cite{Morgan:2004ys,Green:2006cb,Finkbeiner:2009ug}. Here we use
$\langle \cos \gamma_* \rangle$, the WIMP recoil angle with respect to
the Earth's motion. Because of the rotation of the Earth, on average
$\langle \cos \gamma \rangle$ or $\langle \cos \gamma_* \rangle$
should be consistent with 0 for standard backgrounds.

Because XENON100 has excellent spatial resolution, we assume that the
recoil track angle can be determined to 10 degrees. We compute the
exposures required to obtain a $5 \sigma$ result for $\langle \cos
\gamma_* \rangle$ 90$\%$ of the time. We allow for a uniform
(isotropic) background, though the XENON100 background should be
negligibly low. The results are shown in Fig.~\ref{fig:expbench}. The
required exposures roughly correspond to a minimum of 16 events at
zero background.

\begin{figure*}[htpb]
\centering
(a) \begin{minipage}[ct]{.97\linewidth}
      \includegraphics[width=.3\textwidth]{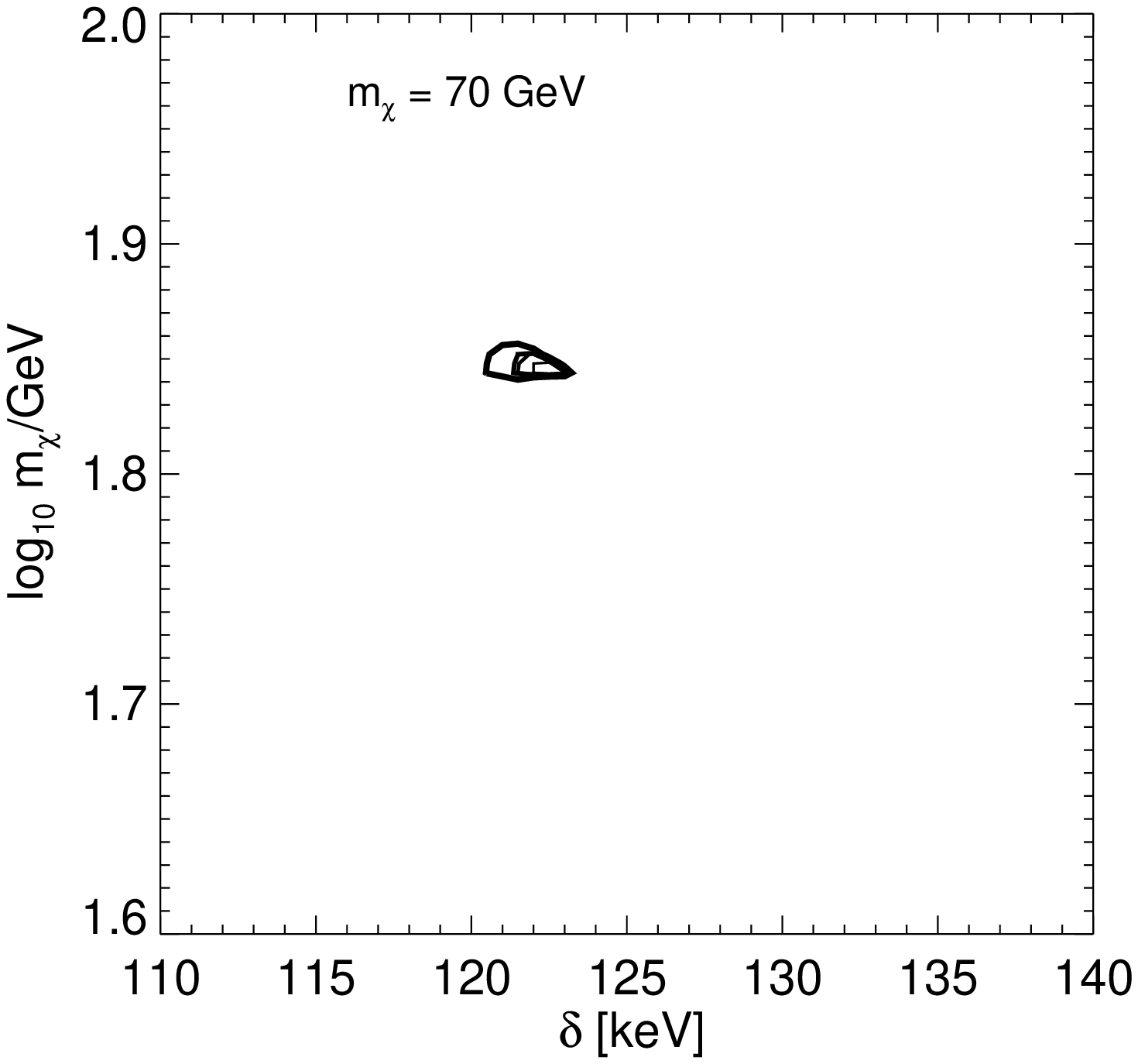}\hskip 0.2in
      \includegraphics[width=.3\textwidth]{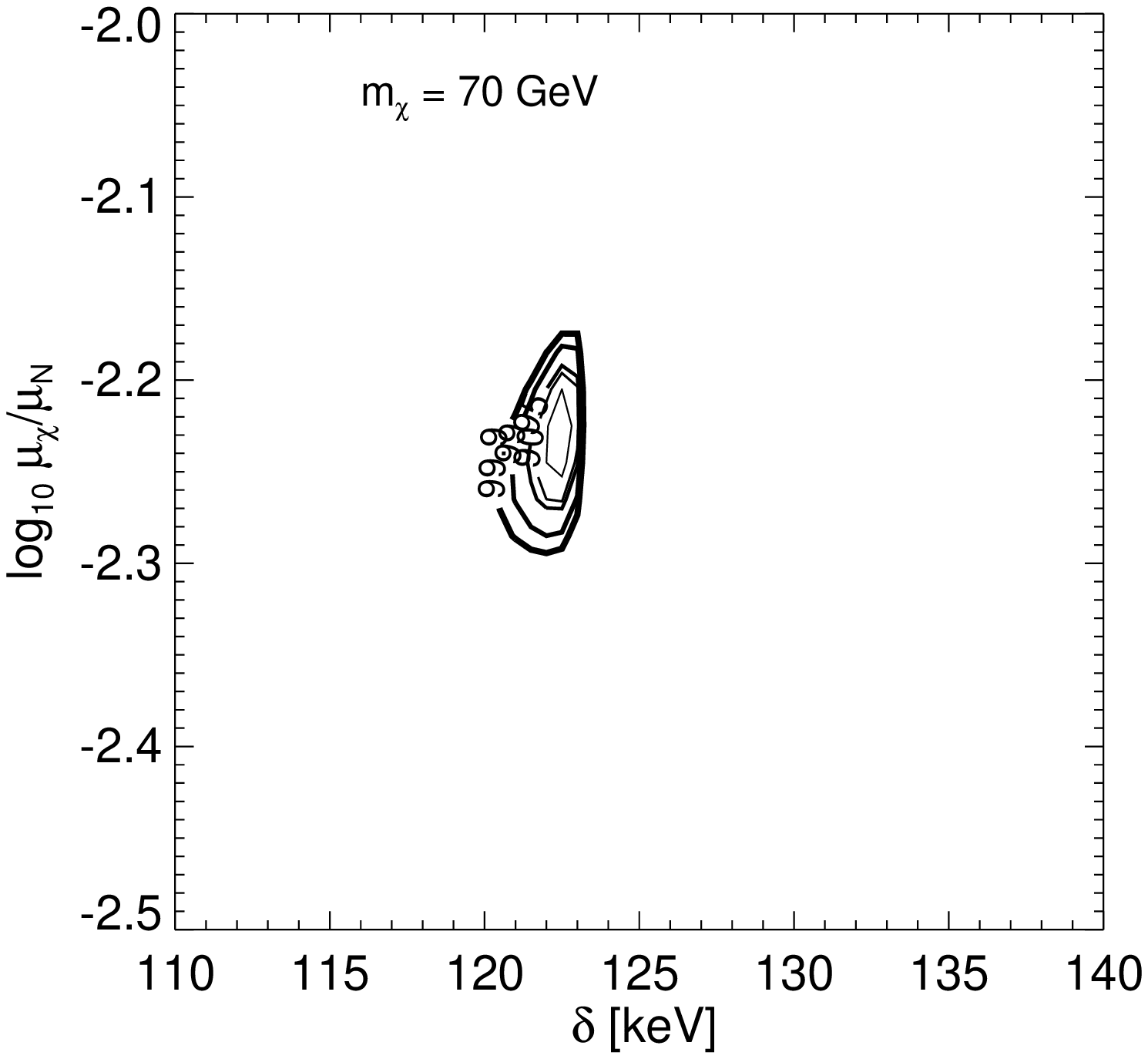}\hskip 0.2in
      \includegraphics[width=.3\textwidth]{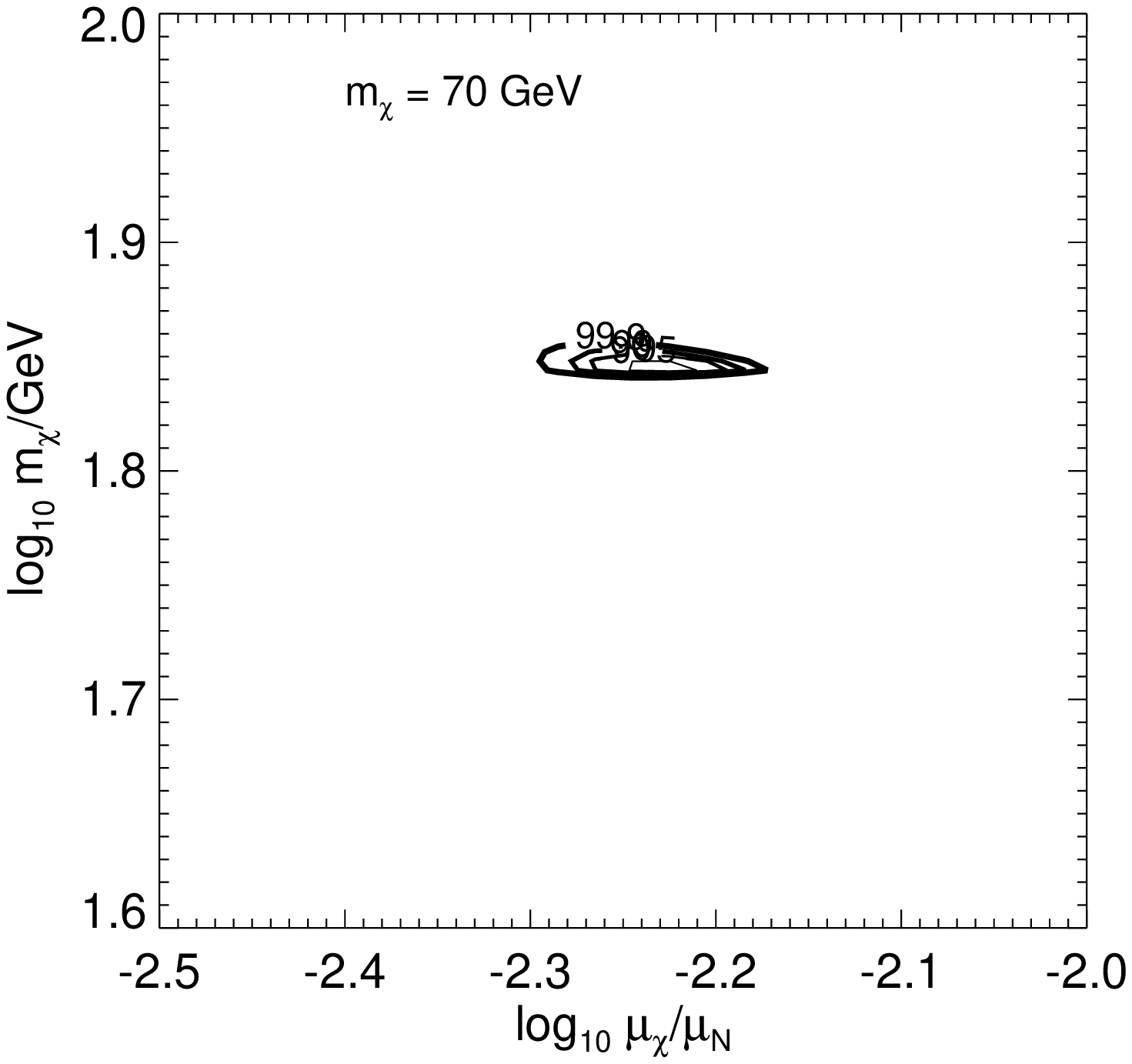}\\
    \end{minipage}
(b) \begin{minipage}[ct]{.97\linewidth}
      \includegraphics[width=.3\textwidth]{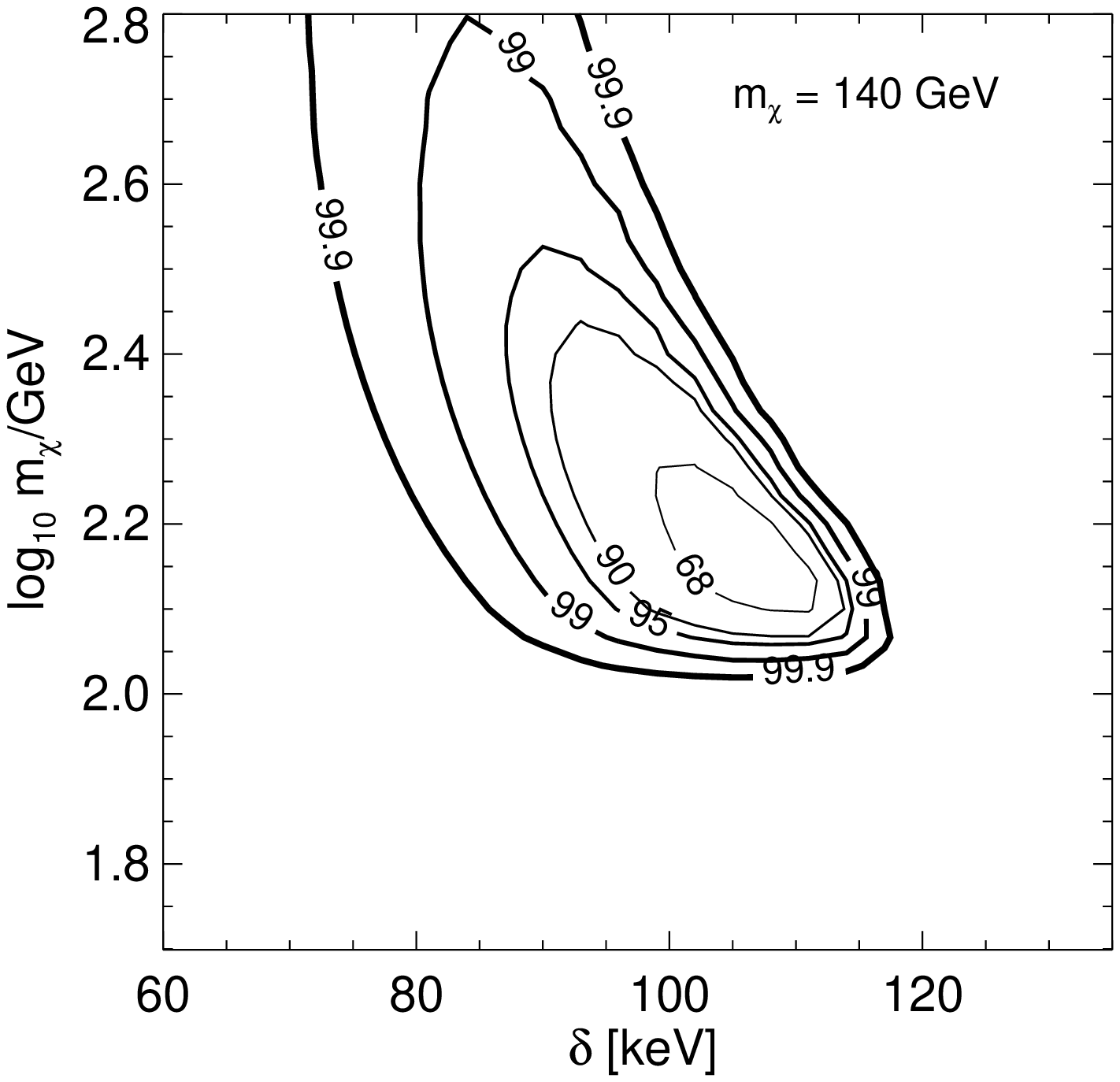}\hskip 0.2in
      \includegraphics[width=.3\textwidth]{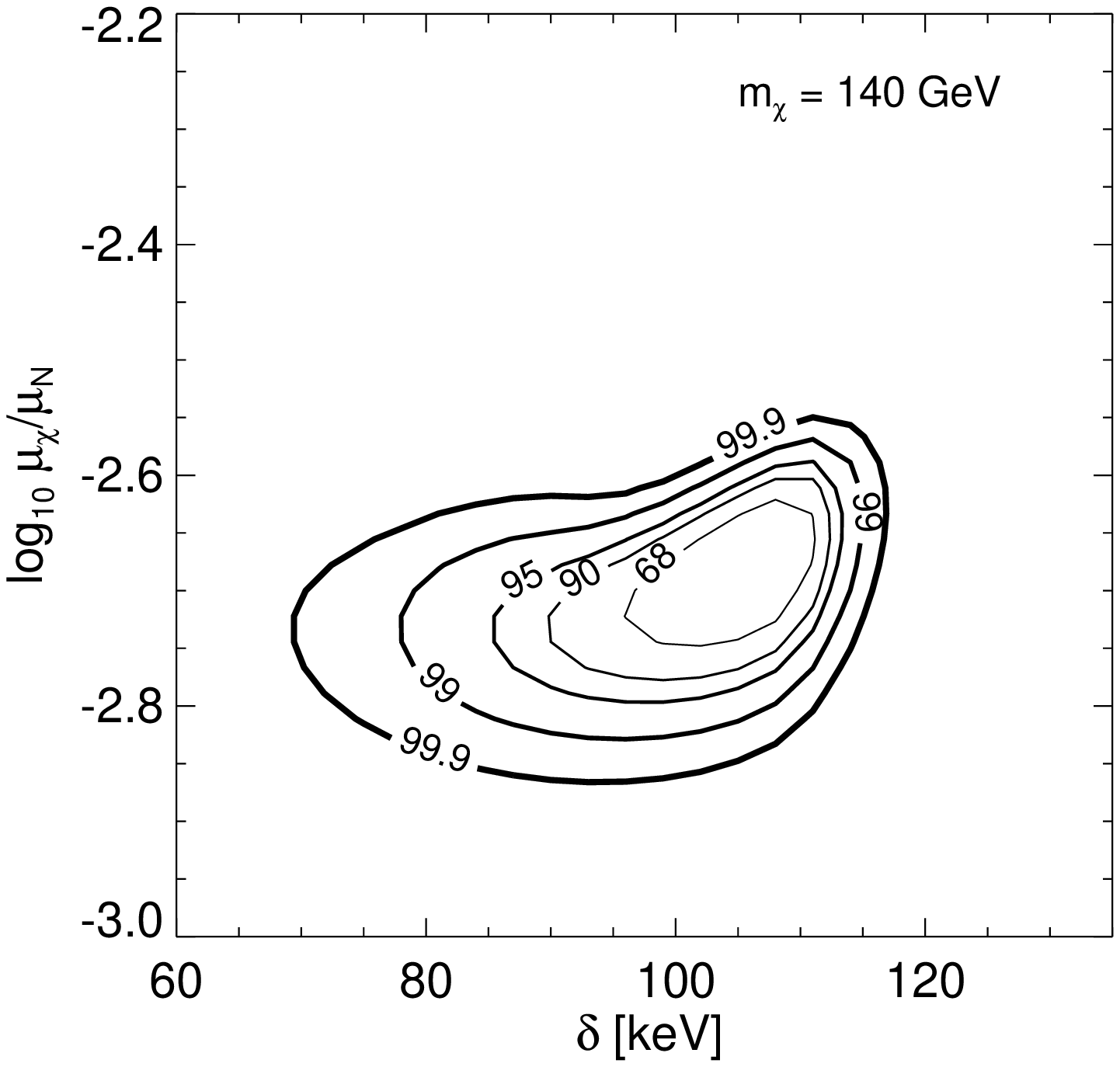}\hskip 0.2in
      \includegraphics[width=.3\textwidth]{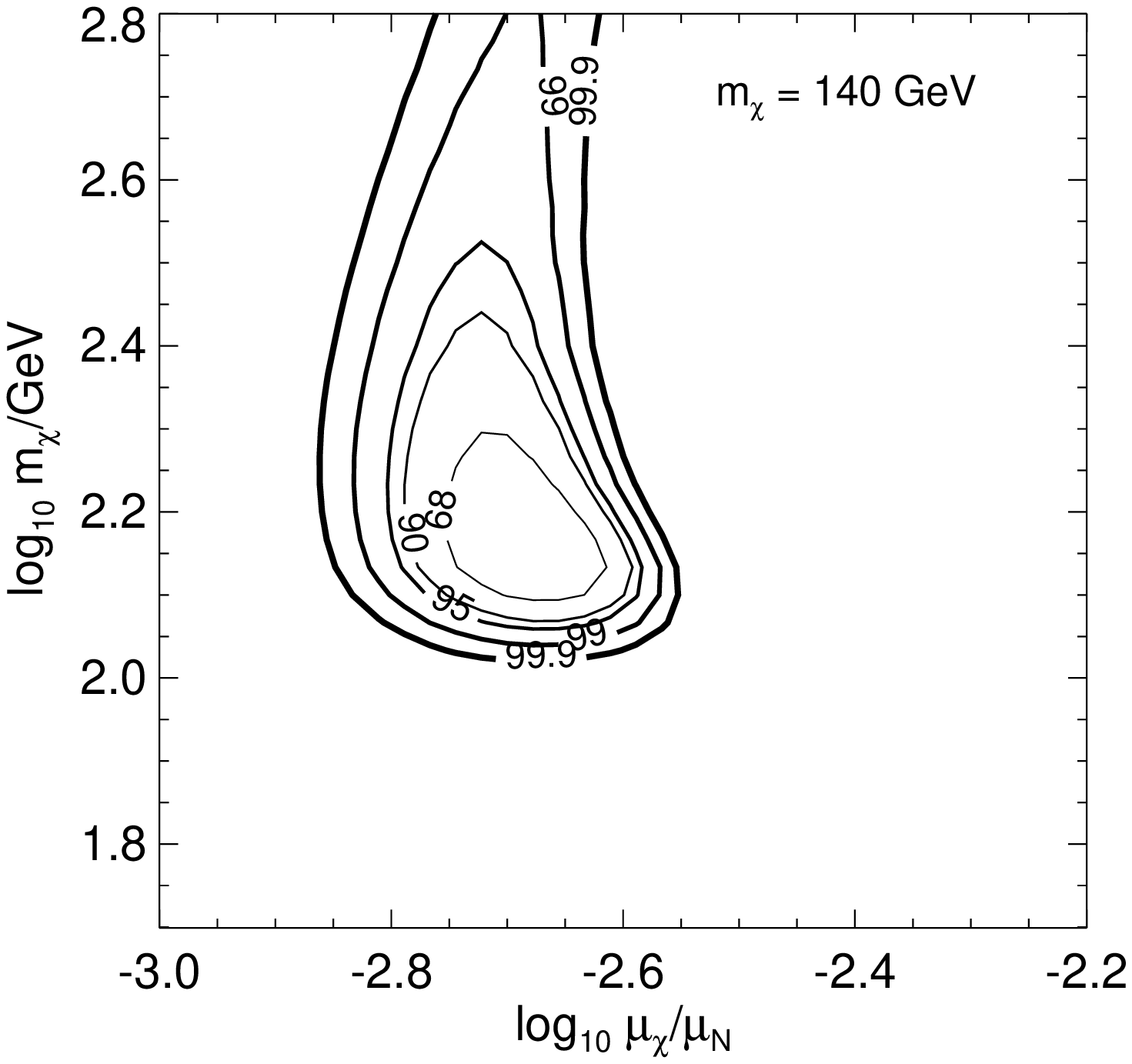}\\
    \end{minipage}
(c) \begin{minipage}[ct]{.97\linewidth}
      \includegraphics[width=.3\textwidth]{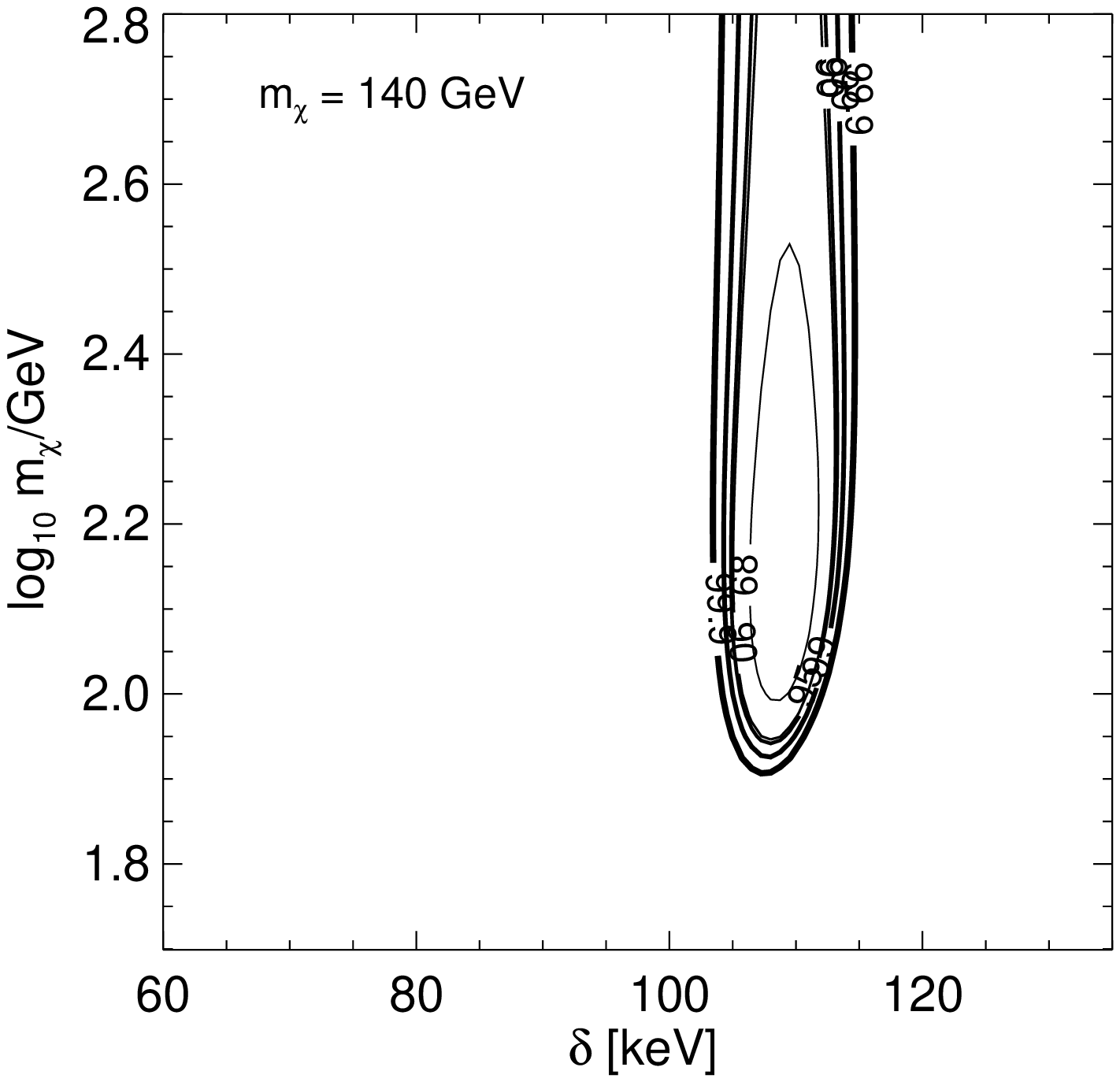}\hskip 0.2in
      \includegraphics[width=.3\textwidth]{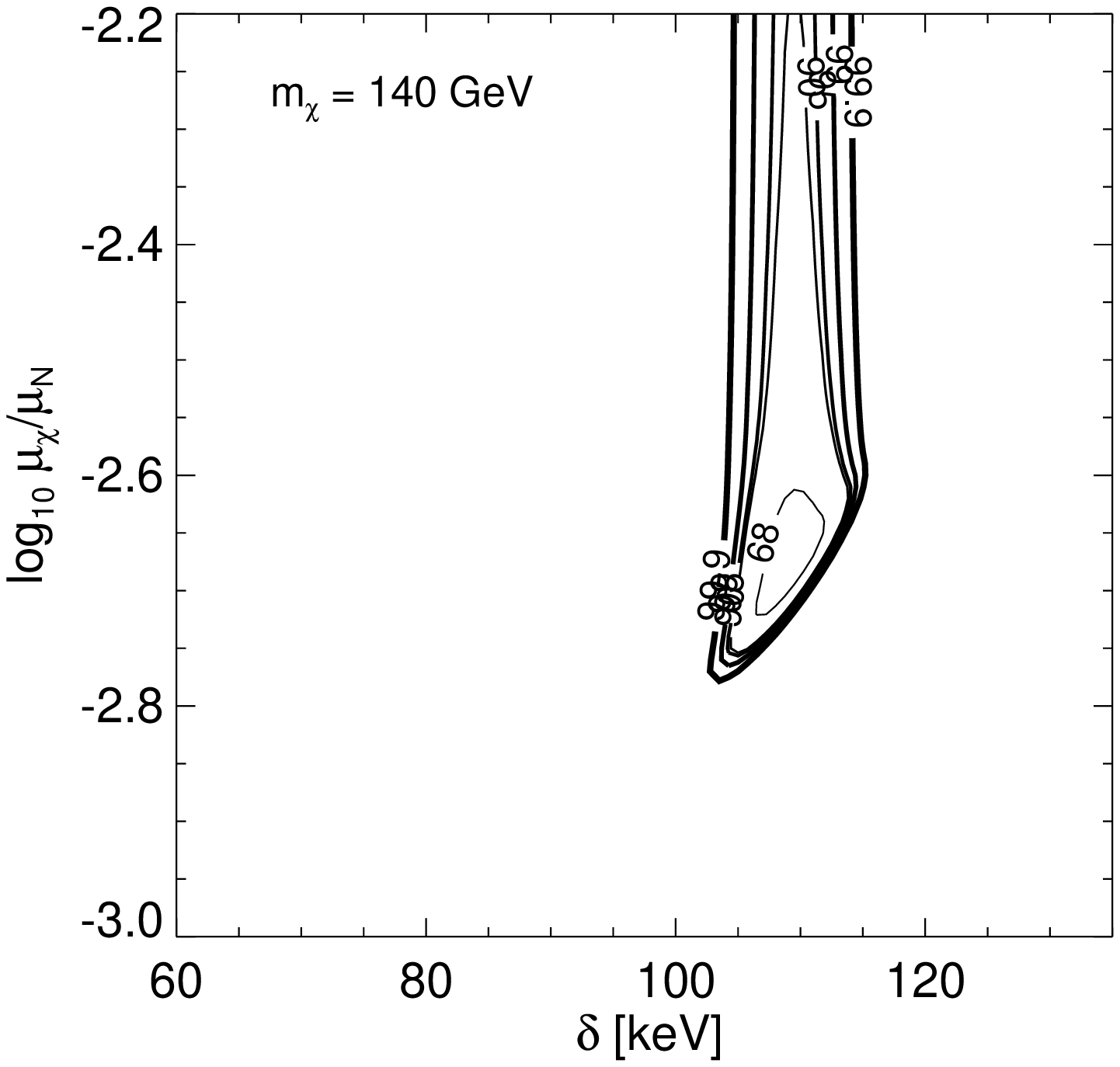}\hskip 0.2in
      \includegraphics[width=.3\textwidth]{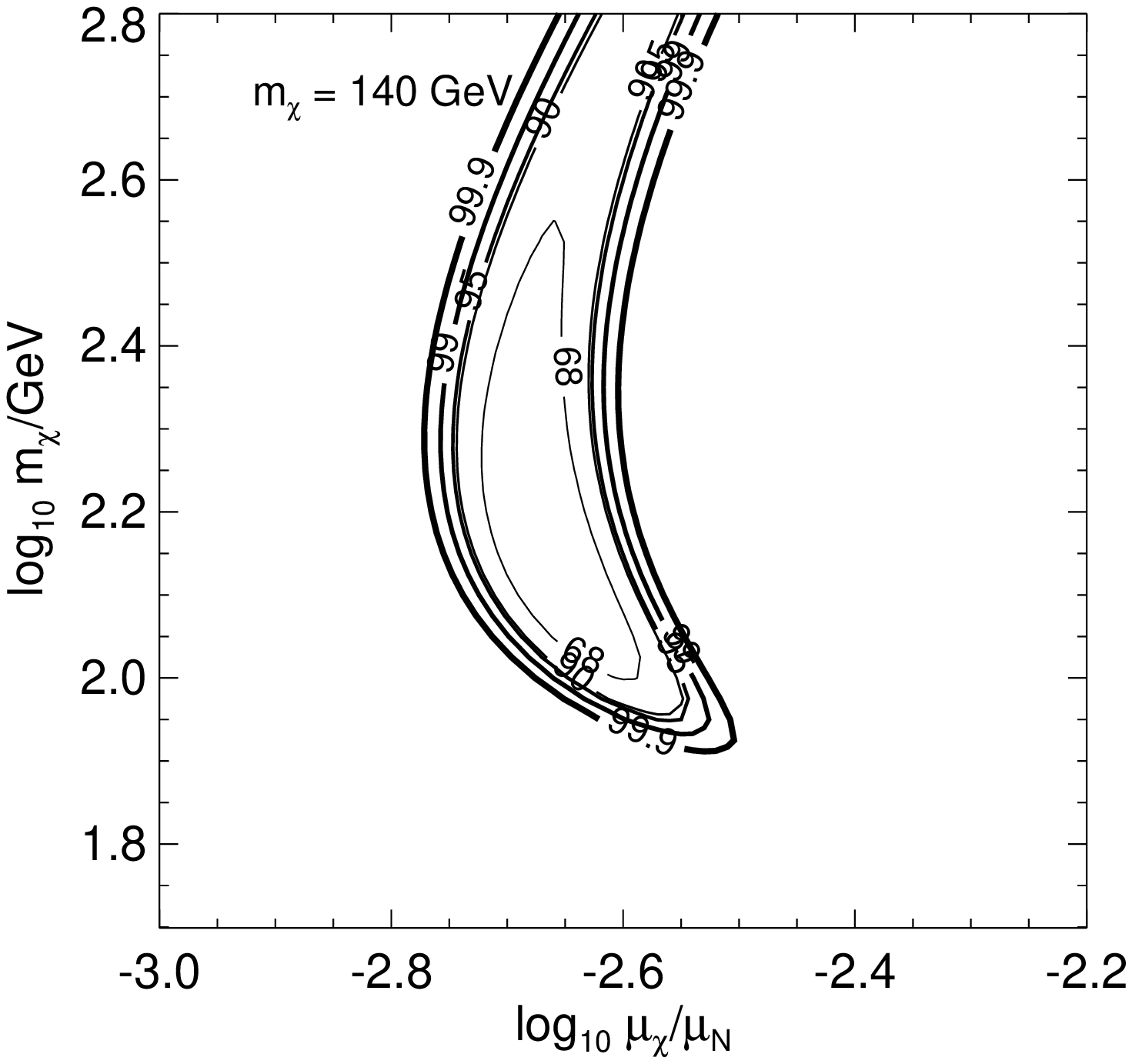}\\
    \end{minipage}
\centering
\caption{Confidence levels for determining $m_\chi$, $\delta$, and
$\mu_\chi$, marginalized over the third parameter for each
two-dimensional slice.  We assume an exposure of 5000 \kgday 
on Xe in the energy range $10-80\kevr$. The plots show sensitivity to
the MiDM parameter space, using only delayed coincidence data, for
the (a) $m_\chi = 70$ GeV benchmark, and (b) $m_\chi = 140$ GeV
benchmark. (c) shows the sensitivity using only nuclear recoil events,
for the $m_\chi = 140$ GeV benchmark. The directional information is a
better test of $m_\chi$ and $\mu_\chi$. The case with $m_\chi = 300$
GeV looks similar to $m_\chi = 140$ GeV.}
\label{fig:lnlbench}
\end{figure*}


\subsection{Parameter Estimation \label{sec:paramestimation}}

The predicted rate for delayed coincidence events at XENON100 is only
a few counts per 1000 kg $\cdot$ day. However, the additional recoil track
information makes it possible to obtain an excellent measurement of
the model parameters.

We perform a likelihood analysis, as described in
\cite{Finkbeiner:2009ug}, over the parameter space of $m_\chi,
\delta$, and $\mu_\chi$. We compute the (relative) log likelihoods for
${\cal E}$ \kgday on Xe, with nuclear recoil energy range
$10-80\kevr$. We neglect the effects of imperfect angular and energy
resolution. (The XENON100 energy resolution is $\sim 10\%$ in this
energy range, and we estimate an angular resolution of 10 degrees.)
The log likelihood is
\begin{equation}
\ln{\cal L}_{tot}(p) ={\cal E} \int dx {\bigg(} \mu(x;p_0) \ln \mu(x;p) - \mu(x;p) {\bigg)} 
\end{equation}
where $p$ refers to $(m_\chi, \delta, \mu_\chi)$ and $x$ refers
generically to the event space of either $E_R$ or
$(E_R,v_*,\cos\gamma_*)$. $p_0$ are the true model parameters. $\mu$ is
the rate for parameters $p$. If there is only nuclear recoil energy
information,
\begin{equation}
  \mu(E_R; p) \equiv \frac{dR}{dE_R}(E_R;p) + dR_{BG},
\end{equation} 
in cpd/kg/keVr for parameters $p$. We assume the background rate,
$dR_{BG}$ is known and negligibly small.

If there is directional information,
\begin{equation}
  \mu(E_R,v_*,x_*; p) \equiv 
     \eta_{.15}(p) \frac{dR}{dE_R dv_* dx_*}(E_R,v_*,x_*;p) 
       + \frac{dR_{BG}}{dv_* dx_*}, 
     \nonumber
\end{equation} 
where $x_* = \cos \gamma_*$.
$\eta_{.15}(p)$ is the efficiency, for parameters $p$, at XENON100.

In Fig.~\ref{fig:lnlbench} we show the sensitivity to MiDM parameters
if (1) only nuclear recoil information is used and (2) if only delayed
coincidence events are considered for 5000 kg $\cdot$ day. We show confidence
levels of (68, 90, 95, 99, and 99.9\%). We neglect the Earth's
velocity about the Sun since a livetime of order a year is needed for
5000 kg~$\cdot$~day.

\begin{figure}[th]
\centering
\includegraphics[width=.45\textwidth]{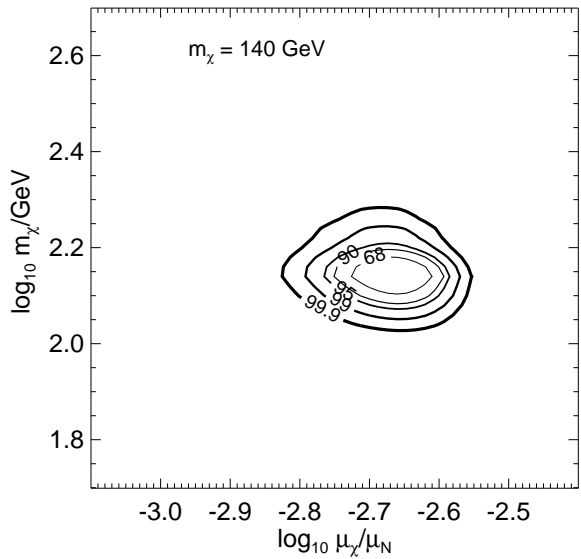}
\centering
\caption{Confidence levels for determining $m_\chi$ and $\mu_\chi$
using WIMP recoil tracks, assuming $\delta$ is already measured from
the photon energies or nuclear recoil spectrum. Here we take the
$m_\chi = 140$ GeV benchmark and assume an exposure of 5000 \kgday on
Xe in the energy range $10-80\kevr$.
\label{fig:lnldelta}}
\end{figure}

Despite the reduction by a factor of 10-50 in events, the directional
data is a much stronger constraint on $m_\chi$ and
$\mu_\chi$. $\delta$ can also be determined from the $E_R$ data or the
photon energies. In Fig.~\ref{fig:lnldelta} we show the sensitivity to
$m_\chi$ and $\mu_\chi$ for the $m_\chi = 140$ GeV benchmark, assuming
that $\delta$ is already known. The directional information breaks the
degeneracy in $m_\chi$ and $\mu_\chi$ when only nuclear recoil
information is used.

\subsection{Measurement of Both Recoils}

So far, we considered measurement of the WIMP recoil velocity vector
from delayed coincidence events. With a gaseous directional detector,
it is also possible to obtain the recoil angle of the nucleus. Then
$m_\chi$ and $\delta$ are highly constrained. For such events there
are 4 equations and 5 unknowns: $m_\chi$, $\delta$, and $\vec
v$. However, one can obtain $\delta$ from the energy peak of the
coincident photons. Then it is possible to measure the WIMP mass and
velocity with just 1 WIMP scattering event. The mass is
determined by the following equation:
\begin{equation}
  m_\chi = \frac{ 2 m_N E_R }{ 2 (\delta + E_R) - \sqrt{2 m_N E_R}\
     \hat q \cdot \vec v_*}.
\end{equation}
Since $\vec q$ and $\vec v_*$ are measured, the initial WIMP velocity
$\vec v$ is then fixed by momentum conservation. A direct measurement
of the WIMP velocity distribution is then also possible.

\section{Conclusions}

The magnetic inelastic dark matter model has an interesting and
previously unstudied signature at direct detection experiments: a
delayed coincident photon with energy $\delta$. Observation of such
photons would also allow current direct detection experiments to
become excellent directional detectors. 

Motivated by the MiDM setup, we studied several benchmark model
parameters that can fit the combined DAMA/NaI and DAMA/LIBRA data.
Given the rapidly improving constraints from other experiments, we
feel that MiDM is currently the best hope for a dark matter
interpretation of DAMA -- \emph{and it predicts a low-background
signature detectable with current experiments.}

With 5000 \kgday of exposure, XENON100 can detect the angular
modulation of the recoils and determine the MiDM model
parameters. While we focused on benchmarks in MiDM, we emphasize that
such a delayed coincidence signal is worth searching for in
general. Such events, if found, carry much more information than
simple nuclear recoils, and would provide more direct access to the
WIMP velocity distribution in our halo. \\

\begin{acknowledgments}
We thank Peter Sorensen, Neal Weiner, and Itay Yavin for useful
discussion. We are partially supported by NASA Theory grant
NNX10AD85G.  This research made use of the NASA Astrophysics Data
System (ADS) and the IDL Astronomy User's Library at
Goddard \footnote{Available at \texttt{http://idlastro.gsfc.nasa.gov}}.
\end{acknowledgments}

\bibliography{midm}

\end{document}